\documentclass[bibyear]{aa}  
\pdfoutput=1
\usepackage{graphicx}
\usepackage{txfonts}
\begin{document} 

\newcommand{\water}{H$_2$O}
\newcommand{\micron}{$\mu$m}
\newcommand{\kms}{km\,s$^{-1}$}
\newcommand{\mdot}{$\dot{\rm{M}}$}
\newcommand{\Myr}{$M_{\sun}$\,yr$^{-1}$}
   \title{A database of circumstellar OH masers}


   \author{D. Engels
          \inst{1}
          \and
          F. Bunzel\inst{1}\fnmsep\thanks{Current address: 
            Max Planck Institute for Meteorology,
            Bundesstraße 53, 20146 Hamburg, Germany.
           }
          }
   \institute{Hamburger Sternwarte, Gojenbergsweg 112,
           21029 Hamburg, Germany\\
              \email{dengels@hs.uni-hamburg.de, felix.bunzel@mpimet.mpg.de}
             }

   \date{Received ???; accepted ???}

 

\abstract{We present a new database of circumstellar OH masers at
  1612, 1665, and 1667 MHz in the Milky Way galaxy. The database
  (version 2.4) contains 13655 observations and 2341 different stars
  detected in at least one transition. Detections at 1612\,MHz are
  considered to be complete until the end of 2014 as long as they were
  published in refereed papers.  Detections of the main lines (1665
  and 1667 MHz) and non-detections in all transitions are included
  only if published after 1983. The database contains flux densities
  and velocities of the two strongest maser peaks, the expansion
  velocity of the shell, and the radial velocity of the star.  Links
  are provided for about 100 stars ($<$5\% of all stars with OH
  masers) to interferometric observations and monitoring programs of
  the maser emission published since their beginnings in the
  1970s. Access to the database is possible over the Web
  (www.hs.uni-hamburg.de/maserdb), allowing cone searches for
  individual sources and lists of sources. A general search is
  possible in selected regions of the sky and by defining ranges of
  flux densities and/or velocities. Alternative ways to access the
  data are via the German Virtual Observatory and the VizieR library
  of astronomical catalogs.  } 

   \maketitle
%

\section{Introduction}
Stars on the asymptotic giant branch (AGB) and red supergiants (RSG)
lose copious amounts of mass and form spherically symmetric
circumstellar envelopes (CSE) where numerous molecular species are
formed. The oxygen-rich AGB stars and RSGs often show maser emission
from OH (at 1612, 1665, and 1667 MHz), \water\ (mostly at 22 GHz), and
SiO (mostly at 43 and 86 GHz) molecules (Elitzur \cite{Elitzur92},
Gray \cite{Gray12}).  Especially among optically obscured AGB stars,
masers easily allow the determination of the chemistry and of the
expansion velocities of their CSEs, and provide radial velocities of
the stars.  Detailed charts of the mass distribution and their
movements can be obtained by interferometric studies (Bains et al.
\cite{Bains03}).  A sizeable fraction of OH/IR stars do not show the
large-amplitude variations commonly seen on the AGB, implying that OH
masers are also sustained  during the early phase of post-AGB evolution
(Engels \cite{Engels02}).  Peculiar maser properties (e.g.  extraordinarily
high velocities or axi-symmetric morphologies) pinpoint  infrared
sources, which are identified as obscured objects in a more advanced
post-AGB evolution phase (Zijlstra et al. \cite{Zijlstra01}).  The
high luminosity of the masers together with the radial velocity
information provided, allow the study of stellar kinematics up to the
Galactic Center region (Sjouwerman et al. \cite{Sjouwerman02}),
and yield  evidence of, for example,  a bar in the Milky Way (Deguchi et
al. \cite{Deguchi02}, Habing et al.  \cite{Habing06}).

New radio observatories like the Atacama Large Millimeter Array (ALMA)
and the forthcoming Square Kilometre Array (SKA) amply enhance the
capabilities for astrophysical research with masers. They will
multiply the number of observable maser transitions and will allow the
study of masers to be extended to nearby galaxies. With the Atacama
Pathfinder Experiment (APEX) the study of a number of \water\ maser
transitions in evolved stars is already in progress (Menten et
al. \cite{Menten08}).  A spectral line survey at 1 mm of the prototype
O-rich star VY~CMa already discovered a new \water\ maser and detected
several SiO masers (Tenenbaum et al. \cite{Tenenbaum10}), showing the
potential for maser research with ALMA in the submm and mm range. The
unprecedented sensitivity of the SKA will allow systematic and
detailed studies of OH and possibly \water\ masers associated with
evolved stars and star forming regions in the Magellanic Clouds and
members of the Local Group (Etoka et al. \cite{Etoka14}). Future
research aiming to compare the new stellar maser detections in
extragalactic systems with the properties of better studied Galactic
masers would benefit from efficient means to access the observations
of the ``classical'' molecular maser transitions made over the past
$\approx$ 40 years.

Since their discovery, several thousand observations to detect masers
in CSEs have been made. For OH masers, catalogs listing detections in
AGB stars and RSGs are more than 20 years old (te Lintel Hekkert et
al. \cite{TeLintel89}; Benson et al.\cite{Benson90}) and are more or
less outdated.  The \mbox{te Lintel Hekkert} et al. catalog contains
detected 1612 MHz OH masers with their flux densities and velocities
and covers the literature until 1984. The Benson et al. catalog lists
references to OH maser observations and covers the years until
1989. The number of detected 1612 OH masers listed are 439 and 713,
respectively.  Since then the number of detected masers has almost
tripled. The last OH maser catalog containing main-line detections is
from Engels (\cite{Engels79}).

\begin{figure*}
\centering
\includegraphics[width=18cm]{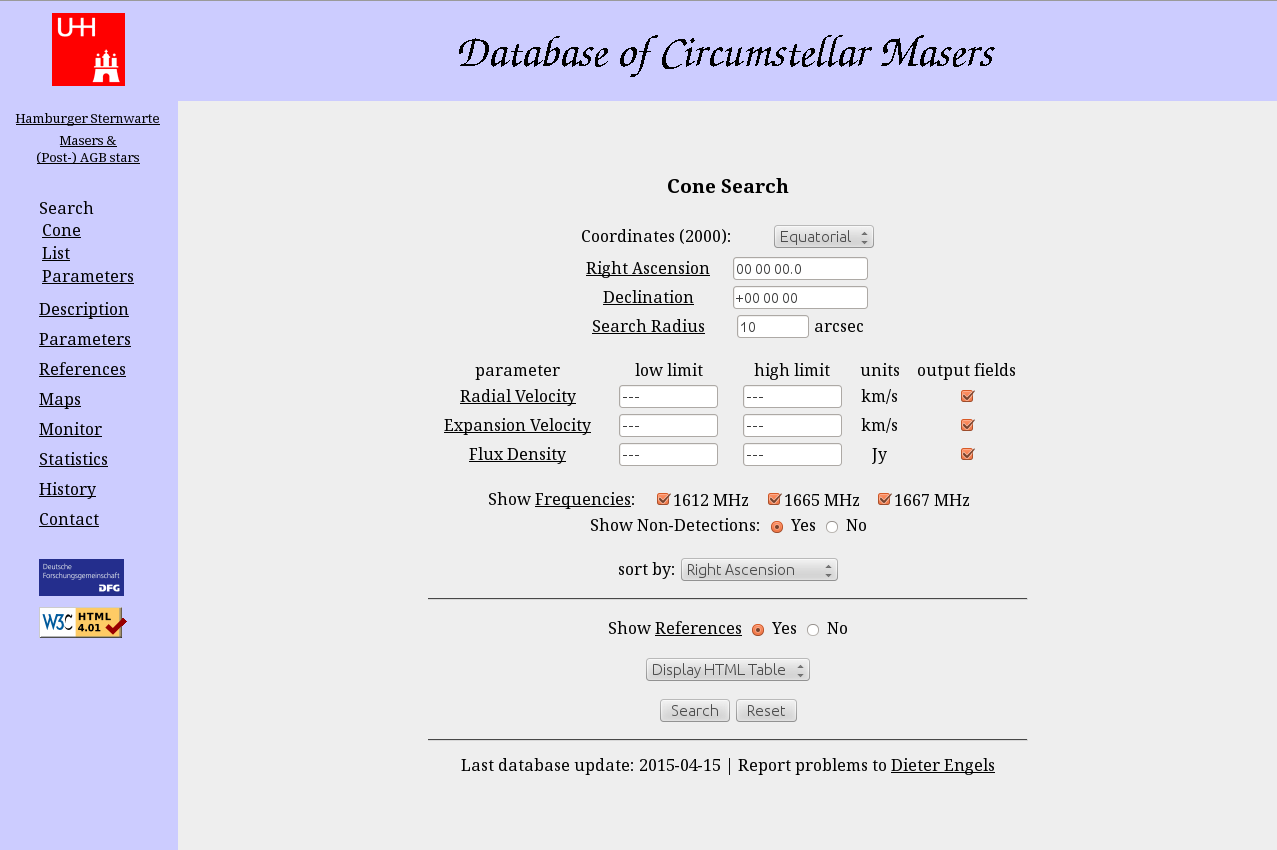}
   \caption{Opening web page for access to the database of OH
masers in circumstellar shells. The address is 
               www.hs.uni-hamburg.de/maserdb             }
              \label{input_mask}%
    \end{figure*}

We have now compiled a database of circumstellar OH masers, which
includes almost all stars in the Milky Way with OH masers ever
detected in one or several of the transitions at 1612, 1665, and 1667
MHz and contains in addition published observations with
non-detections in these transitions since 1984. A first version of the
database was published in 2007 by Engels \& Bunzel (\cite{Engels07b})
and has been updated continuously since then. An extension that will
also cover \water\ and SiO masers is in the works.

In the last decade no larger surveys for OH masers have been made, and
research has shifted toward interferometric and variability studies of
known stellar masers. The interferometric studies focus, for example,
on astrometry to derive distances (Vlemmings \& van Langevelde
\cite{Vlemmings07}) or on polarization properties to derive magnetic
field strengths (Amiri et al. \cite{Amiri11}).  To take this
development into account, these follow-up observations were included
as separate tables in the more recent versions of the database.

The database can be accessed via an internet portal
(Fig. \ref{input_mask}), which allows an efficient search using a
variety of parameters.  The database is also integrated in the Virtual
Observatory\footnote{The IVOA identifier is
  ivo://org.gavo.dc/ohmaser/q/scs} and is available through the Cone
Search protocol for simple positional queries\footnote{Protocol access
  URL: http://dc.g-vo.org/ohmaser/q/scs/scs.xml} as well as the Table
Access protocol for advanced database queries\footnote{Protocol access
  URL: http://dc.g-vo.org/tap} allowing complex conditions and joins
with other, potentially user-provided, catalogs. In tabular format the
database content is available at the VizieR library of astronomical
catalogs\footnote{The tables are available in electronic form at the CDS via
  anonymous ftp to cdsarc.u-strasbg.fr (130.79.128.5) or via
  http://cdsweb.u-strasbg.fr/cgi-bin/qcat?J/A+A/}.

The content of the database is described in Sect. \ref{content}. The
access of the database and the different selection options to search
in it are described in Sect. \ref{access_base_main}. In Section
\ref{properties} we analyze the properties of OH masers in CSEs using
the content of the database and we finish the paper by summarizing
the types of stars hosting OH maser emission and the role of the
database for future maser research.

\section{Contents of the OH maser database \label{content}}
The database is made up of a main table containing  the survey
observations and two auxiliary tables containing monitoring and
interferometric follow-up observations for detected masers. The survey
observations were made with single-dish radio telescopes and beginning
in the 1980s also with the Very Large Array (VLA) and the Australian
Telescope Compact Array (ATCA) interferometers.

\subsection{Setup of the tables and completeness}
The main table was set up with the contents of the te Lintel Hekkert
et al. catalog. The following search for OH maser observations was
made in the literature (refereed papers) published between 1984 and
2014. The database is considered to be almost complete for stars with
1612 MHz detections, but contains no stars observed and published
prior to 1984 without detected 1612 MHz masers, while for the main
lines (1665 and 1667 MHz) the database contains only measurements
published after 1984. The database is restricted to Galactic masers.
The auxiliary tables contain publications from 1974 onward,
covering interferometric and monitoring observations since their
beginnings.

\subsection{Information content of the main table}
In each publication containing circumstellar OH maser observations we
searched for the following information:
\begin{itemize}
\item the names of the stars or the OH maser designation; 
\item the transition frequency 1612, 1665, or 1667 MHz;
\item coordinates;
\item peak flux densities and velocities of maser features; 
\item radial velocity $v_{rad}$ of the star; 
\item expansion velocity $v_{exp}$ of the CSE.
\end{itemize}

Not all information is always found and sometimes the meaning of a
particular parameter or its unit differs between papers. We then
transformed values and units to a common definition. All such
modifications of the original data are logged in special reference
pages (see Sect. \ref{ref_pages}), which are maintained for
each paper included in the database.

Each database entry contains several fields used for administrational
purposes. The ``measurement number'' uniquely identifies the
particular entry, the ``paper identification number'' identifies the
paper from which the data originates, and the ``source number''
assures that measurements from the same source are linked, even if
names or coordinates differ (the need for such a link is described in
Sect. \ref{access_base}).  The fields ``Mon.'' (for monitoring)
and ``Maps'' are flags, which are logically set to ``true'' if there
is data in the auxiliary tables available.

\subsubsection{Flux densities and velocities catalogued}
At 1612 MHz, the profiles of circumstellar OH masers usually show a
characteristic double-peaked profile, in which the two peaks are
typically separated by $\Delta v \approx 30$ \kms. Peak flux densities
and velocities of these two peaks are usually tabulated in the
discovery papers to characterize the maser.  The double-peaked profile
is created by radially masing regions in the front and  the back of the star, which are moving away from the central star with a
velocity $\Delta v /2$. The database contains therefore the flux
density $f_{blue}$ and the velocity $v_{blue}$ of the (most) blue
shifted and $f_{red}$, $v_{red}$ of the (most) red shifted maser
peak. In a few cases where peak flux densities and/or velocities were
not given this information was taken directly from the published
spectra.  Based on the assumption of a radially symmetric CSE, $\Delta
v /2$ is its expansion velocity $v_{exp}$, while the center of the
velocity interval covered by the two maser peaks is the radial
velocity $v_{rad}$ of the star.  The velocities $v_{exp}$ and
$v_{rad}$ are included in the database.  They are taken either from
the paper or are calculated from the peak velocities.

A different method of selecting the two maser features to be included in
the database and to determine $v_{exp}$ and $v_{rad}$ was used for the
main lines in those cases where the maser had no simple double-peaked
profile. Circumstellar masers in general and the OH maser main features
in particular often consist of many individual maser features spread over
a velocity interval of several \kms. In such cases papers only list peak flux
densities and velocities of the strongest features. Because of the
variability of the masers a particular feature selected, however, only
temporarily characterizes a given maser. Therefore, in principle the
integrated flux and the velocity interval over which maser emission
has been observed are more suitable parameters to characterize a
particular maser. Owing to the lack of information on these parameters
in most papers,  for the OH main features we listed the peak flux
densities and velocities of the two strongest maser features to
characterize the strength of the maser. The velocities $v_{exp}$ and
$v_{rad}$ were then not calculated from these peak velocities, but were
explicitly calculated from the velocity interval covered by the
outermost maser peaks listed.

\subsubsection{Special cases and verification}
If a paper lists several observations of the same star, we
transferred the observation with the strongest peak flux density
measurement to the database. Sometimes only a single maser feature was
observed. The database contains an entry ``spectral type'' that is
set by default to ``D'' (double), but for these observations is set to
``S'' single.  For a few sources the ``spectral type'' is set to
``I'' (irregular) if the original paper gives this classification.
Flux density and velocity for single maser features are listed as $f_{blue}$ and
$v_{blue}$, and $v_{rad} = v_{blue}$ is set to ensure that the source
can be included in searches restricting the radial velocity. The
feature is listed as $f_{red}$ and $v_{red}$, and $v_{rad} = v_{red}$
is set if this can be inferred from other maser observations. For
observations with non-detections the entry $f_{blue}$ contains the
3$\sigma$ value of their sensitivity.

To verify the correct registration of the data in the database, the 
entries in the database and the entries in the paper were compared
manually for two observations randomly taken. However, erroneous
entries cannot be excluded, especially if input tables were not
available in electronic form and data were transferred manually or
from scanned tables.

\begin{figure}
\centering
\includegraphics[width=8.5cm]{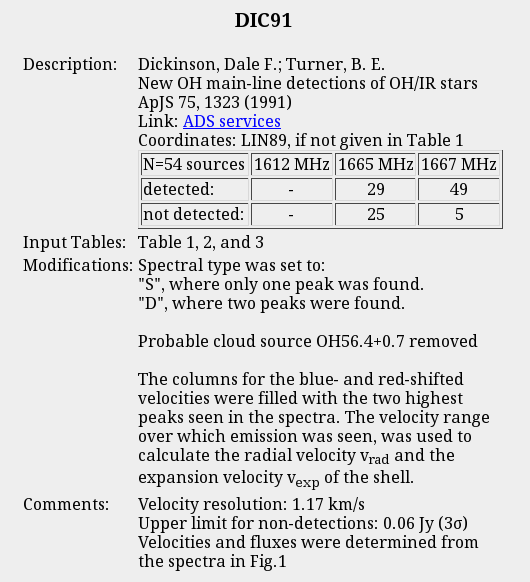}
   \caption{Individual reference page for the main line observations
     of Dickinson \& Turner (\cite{Dickinson91}). Statistics, an
     external link to the SAO/NASA Astrophysics Data System (ADS),
     references for additional data, and modifications to the
       original data in the paper are given.  
              \label{refer_page}%
           }
    \end{figure}

\subsection{Information content of the auxiliary tables}
The entries in the main table are flagged (for details see
Sect. \ref{aux_access}) if monitoring or interferometric follow-up
observations were made for a particular frequency. Information about
such follow-up observations is stored in two auxiliary tables. The
basic entries for the auxiliary tables are
\begin{itemize}
\item the names of the stars or the OH maser designation; 
\item the coordinates;  
\item the transition frequency: 1612, 1665, or 1667 MHz.
\end{itemize}

\begin{figure*}
\centering
\includegraphics[width=18cm]{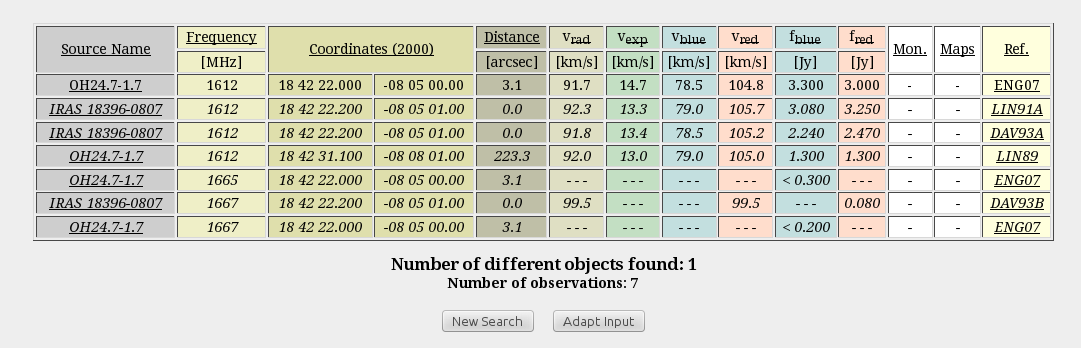}
   \caption{Result from a cone search at the position 18 42 22.2 -08
     05 01 (2000) with a search radius of 10\arcsec. The output page
     displays the content of the main table. The source names are
     linked to the SIMBAD database. Access to entries in the auxiliary
     tables (in this case no entries are available) is provided by the
     links in columns ``Mon.'' and ``Maps''.  We note the inclusion of
     the entry from the te Lintel Hekkert et al. (1989) catalog
     (LIN89).  Although the catalog position is outside the search
     radius, the observation is listed because of an internal link to
     the ``primary observation'' of Engels \& Jim\'{e}nez-Esteban
     (\cite{Engels07}) labeled as ENG07 (see
     Sect. \ref{access_base}).}
              \label{output_mask}%
\end{figure*}

The specific entries for the first auxiliary table, the monitoring
table, are
\begin{itemize}
\item the time interval covered by the observing epochs;
\item if available, the range of flux densities observed.
\end{itemize}

The specific entries for the second auxiliary table, the
interferometry table, are
\begin{itemize}
\item  the name of the interferometer used;
\item the angular resolution;  
\item the sensitivity of the observations;
\item the presence of polarization information indicated by a
yes/no entry; 
\item the number of maps;  
\item the range of observing epochs. 
\end{itemize}

As was done for the main table, special reference pages (see below) are
maintained for each paper included in the database. All modifications
of original data are logged in these pages. As before, the correctness
of the table entries was verified for each paper by a manual
comparison of two random database entries with the original paper.

\subsection{Reference pages  \label{ref_pages}}
In principle, all data were taken directly from the paper as
published. If data from other sources were included (for example
coordinates), or the data was transformed, this has been logged on an
individual ``reference page''. An example is given in
Fig. \ref{refer_page}. A reference page was created for each
paper contained in the database. This page gives the journal, where
the paper was published, a link to the SAO/NASA Astrophysics Data
System\footnote{http://www.adsabs.harvard.edu/.} (ADS), the source of
the coordinates (if not from the paper itself), the number of sources
observed and detected, the source for the input data (e.g. tables of
the printed version or online tables), modifications applied to the
input data, and comments. Thus, the origin of all information included
in the database, and which was not provided directly by the paper,
can be traced back by consulting the reference pages.

\begin{figure*}
\centering
\includegraphics[width=18cm]{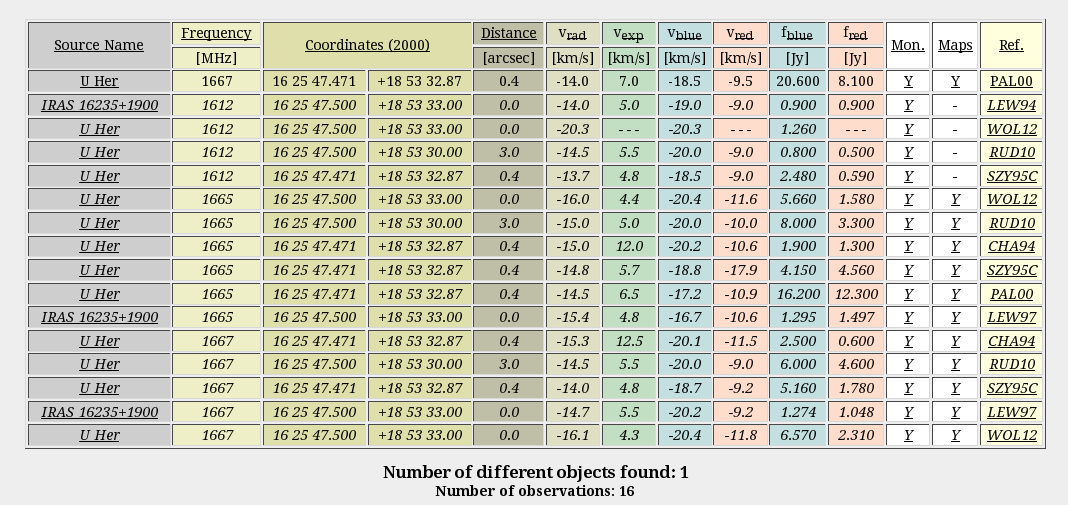}
\includegraphics[width=18cm]{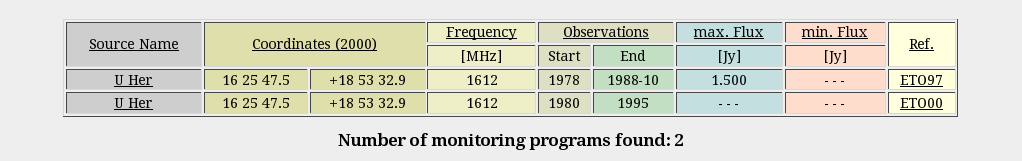}
\includegraphics[width=18cm]{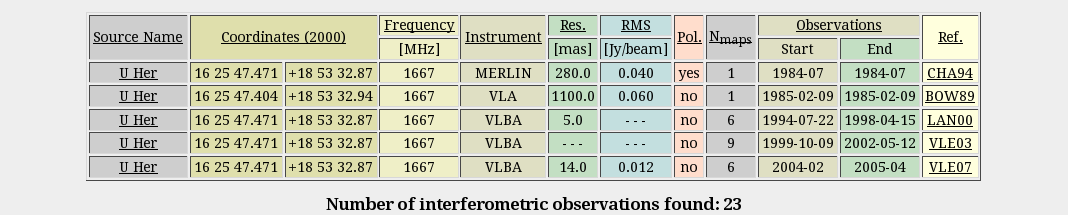}
   \caption{Examples for the content in the main and auxiliary tables
     of the database for the Mira variable U\,Her. {\em Top:} Result
     from the search in the main table. The columns ``Mon.'' and
     ``Maps'' are activated (Y=``Yes'')
 for several frequencies, linking to
     monitoring and interferometric observations at these frequencies
     compiled in the auxiliary tables.  {\em Middle:} Result from the
     search in the monitoring table at 1612 MHz. A maximum
     flux density of 1.5 Jy was reported for U Her during $\approx$11
     years of monitoring.  {\em Bottom:} Compiled
     interferometric observations at 1667 MHz.  }
              \label{UHer}%
    \end{figure*}

\section{Access to the OH maser database \label{access_base_main}}
The OH maser database can be accessed via  the web interface
www.hs.uni-hamburg.de/maserdb (Fig. \ref{input_mask}). The web
interface consists of the main search page and supplementary  pages for
documentation and additional information. 

\subsection{Main search page  \label{access_base}}
On the main page there is a choice of three different search modes:
\begin{itemize}
\item Cone search\\ This mode allows the search for catalog entries
  in the vicinity of a celestial position. Equatorial or Galactic
  coordinate systems can be selected to enter coordinates. The search
  radius is restricted to \mbox{$<1000$\arcsec}. The search can be
  restricted further by limiting the range of frequencies,  radial
  velocities $v_{rad}$,  expansion velocities $v_{exp}$, and  flux
  densities. 

\item List search \\
This mode works in the same way as the cone search mode, but a table with multiple 
search positions can be uploaded.

\item Parameter search\\
This mode allows a general search in the database. It works like
the two previous modes, except that a search area on the sky 
can be defined by coordinates. Without restriction of any of the
parameters the complete content of the database is accessed.
\end{itemize}

Typical output pages for the three search modes are shown in
Figs. \ref{output_mask} and \ref{UHer}. In addition to columns
containing the observation specific information (source name,
frequency, coordinates, distance to the search position, velocities,
and flux densities), an output page contains columns ``Mon.'', ``Maps'', and ``Ref.'' giving links to observations in the auxiliary
tables and to the reference page.

There are several options to restrict the output. The minimum output
is given by source name, frequency, and the links to the auxiliary
tables.  Coordinates, flux densities and velocities, and literature
references for the observations can be added to the output.  It is
possible to suppress the listing of non-detections.  The output can be
sorted by several parameters and can be downloaded as a formatted
ASCII file for off-line analysis.

Cone and list searches require a search radius, and therefore the
completeness of such searches is not given. The reason are the often
quite poor (radio) coordinates given in the earlier literature, which
may deviate from the true position by several arc minutes. Only part of
the observations are listed, especially if small radii are chosen. An
increase in the radius often leads to the inclusion of possibly
non-related neighboring sources into the output. We decided, therefore,
to implement a special mechanism to ensure that all observations for a
particular source are actually listed in the output.  For each OH
maser source a particular observation with accurate coordinates is
chosen as the ``primary observation'' and all other observations of this
source are linked to it as ``secondary observations''.  The link was
implemented in the database by assigning the same internal ``source
number'' to all observations. Observations with largely deviating
coordinates (e.g. $30-300$\arcsec) were assigned to the same source
(i.e. they were linked together) if the velocities of the maser features
coincided.  An example is given in Fig. \ref{output_mask}. All
searches are made among the primary observations, but the result
of the search also contains  all secondary observations. This
optimizes the completeness for position based searches at the expense
of the completeness for searches using the other parameters. This is
the case if, for example, the search is restricted to a minimum flux
density. Sources matching this limit in a secondary observation
but not in the primary observation will not be selected.

More details can be found in the supplementary  pages.

\subsection{Supplementary pages}
The supplementary pages can also be accessed  over the web interface (see
Fig. \ref{input_mask}) and contain information on the organization of
the database (Link: ``Description''), on the definition of the input
and output parameters (Link: ``Parameters''), on the origin of the
data (Link: ``References''), on basic statistics of the content of the
database (Link: ``Statistics''), and on the history of releases (Link:
``History''). Also, the auxiliary tables (Links: ``Maps'' and ``Monitor'') 
can be directly accessed  via the web interface.

The ``Description'' page contains the basic information about the
database as described in this paper and will log future changes to
content and structure. The outdated Description pages from
earlier releases can be accessed by the ``History'' page, and allow  the development of the database to
be followed. The ``Parameters'' page
describes in detail the parameters used (definition, ranges of values,
units).  The link ``References'' lists all papers used for
the compilation of the database. The individual reference pages
(see Sect. \ref{ref_pages}) can be accessed here by a link labeled by an
acronym (for example ENG07 for Engels \& Jim\'{e}nez-Esteban
\cite{Engels07}).  This acronym is also used as a label for the link
in the database output, pointing to the original paper from which a particular
observation came from. The  ``Statistics'' page gives a compressed
overview on the content of the current release of the database and is 
described in Sect. \ref{properties}.

\begin{figure}
   \includegraphics[width=6.0cm,angle=-90]{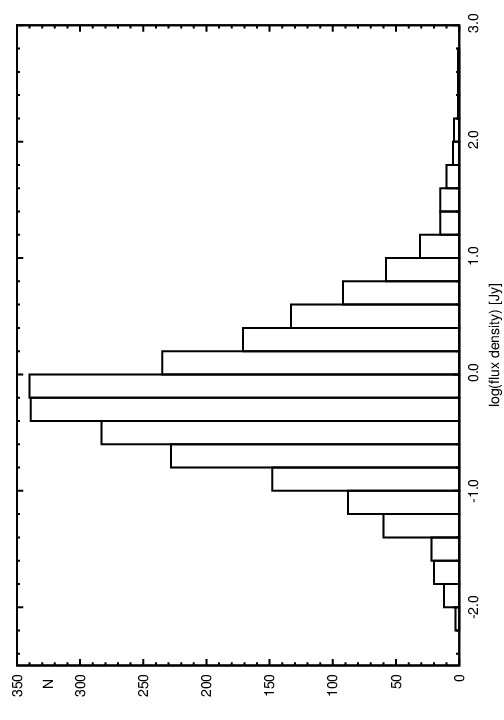}
   \includegraphics[width=6.0cm,angle=-90]{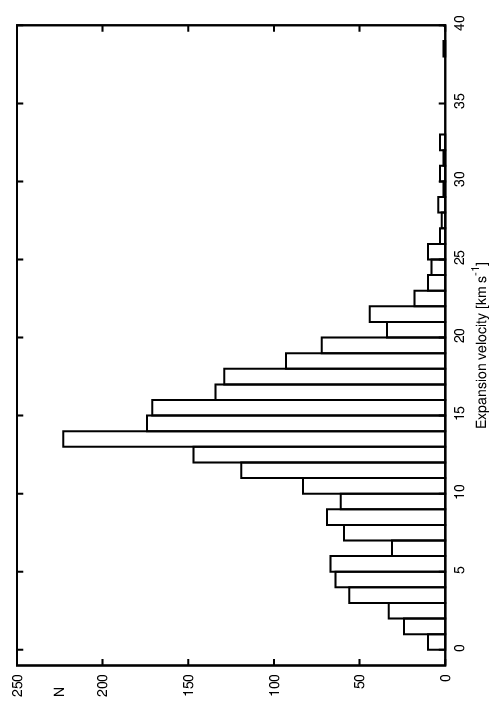}
 \caption{Distribution of the 1612 MHz peak flux densities (upper
   panel) and expansion velocities (lower panel). The maxima are at
   $f_{peak} \approx 0.5$ Jy and $v_{exp} \approx 14$
   \kms. About a dozen sources with $v_{exp} = 40-80$ \kms\ are 
   excluded from the plot.}
 \label{velexp-flux}%
\end{figure}

\subsection{Access to monitoring and interferometric observations \label{aux_access}}
Figure \ref{UHer} displays part of the content of the database for the
Mira variable U\,Her. All OH maser transitions observed in this star were
included in several monitoring programs and interferometric studies.
Therefore links are provided in the columns ``Mon.'' and ``Maps'' for
these frequencies. The 1612 MHz maser was monitored for almost  a decade
as shown in the middle panel of
Fig. \ref{UHer}.  The interferometric observations compiled for the
frequency 1667 MHz are shown in the bottom panel of the same figure.

The full content of the monitoring and interferometric tables can be
accessed directly using the supplementary pages ``Monitor'' and ``Maps''.
The monitoring table contains 180 long-term observations of 78 stars
published in 11 papers.  The interferometric table contains 427
observations of 102 stars published in 66 papers. So far, less than 5\% of
all stars showing OH maser emission have been monitored and/or mapped.

\subsection{Links to external databases}
Links to external databases are provided for SIMBAD\footnote{SIMBAD
  Astronomical Database, Univ. of Strasbourg,\\
  http://simbad.u-strasbg.fr/simbad/.} and for ADS. The link
to SIMBAD for each source in the database is provided in the field
``Source Name''. As the search in SIMBAD is made by the source name a
failure of matches is possible. The link to ADS is used to access the
original literature providing the data.  Access to the literature is
either possible from the individual reference pages
(see Sect. \ref{ref_pages}) or by the label listed in column ``Ref.''
of the output pages.

\section{Properties of OH masers in circumstellar shells \label{properties}}
\subsection{Detection rates}
The database in its version 2.4 contains 13655 OH maser observations
at frequencies 1612 (62\%), 1665 (17\%), and 1667 MHz (21\%). These
observations belong to 6356 different stars with 2341 (37\%) stars
detected in at least one transition.  The overall detection rate is
compatible with the numbers in the catalog of Benson et al.
(\cite{Benson90}), who reported 32\% detections among 2253 stars
observed.  The vast majority (94\%) of the detected sources show
maser emission at 1612 MHz.

\begin{figure*}
\begin{minipage}[t]{18cm}
 \begin{minipage}[t]{8.5cm}
  \begin{flushleft}
   \includegraphics[width=6.5cm,angle=-90]{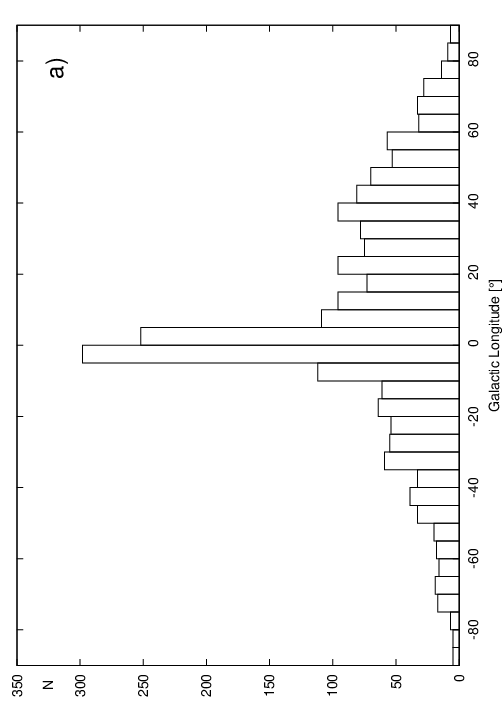}
  \end{flushleft}
 \end{minipage}
 \hfill
 \begin{minipage}[t]{8.5cm}
  \begin{flushright}
   \includegraphics[width=6.5cm,angle=-90]{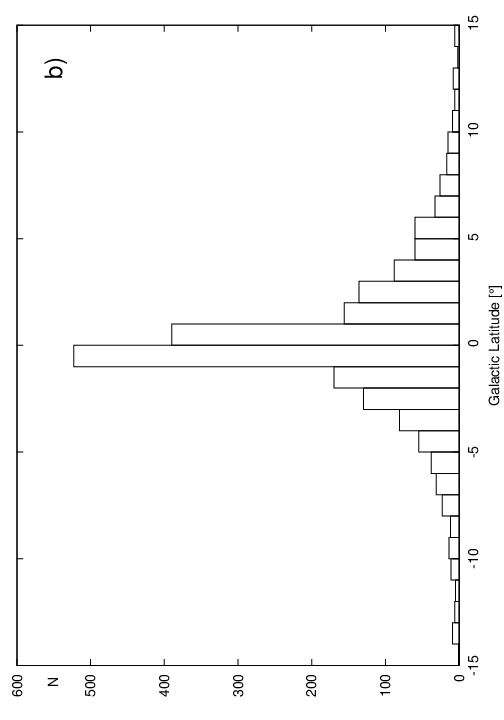}
  \end{flushright}
 \end{minipage}
\hfill
 \begin{minipage}[t]{8.5cm}
  \begin{flushleft}
   \includegraphics[width=6.5cm,angle=-90]{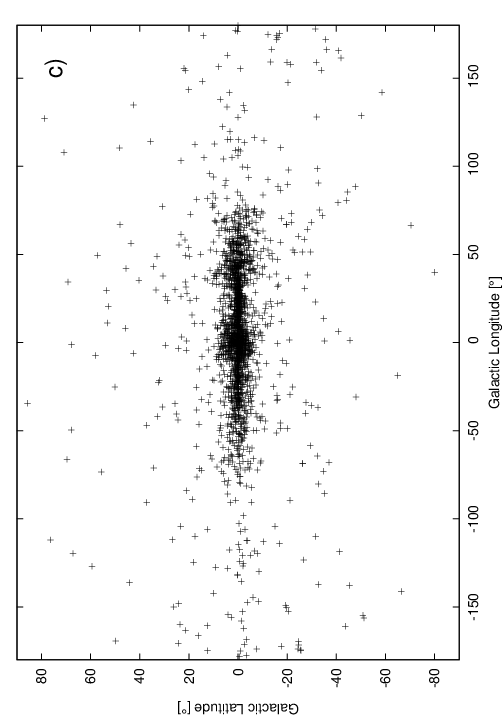}
  \end{flushleft}
 \end{minipage}
 \hfill
 \begin{minipage}[t]{8.5cm}
  \begin{flushright}
   \includegraphics[width=6.5cm,angle=-90]{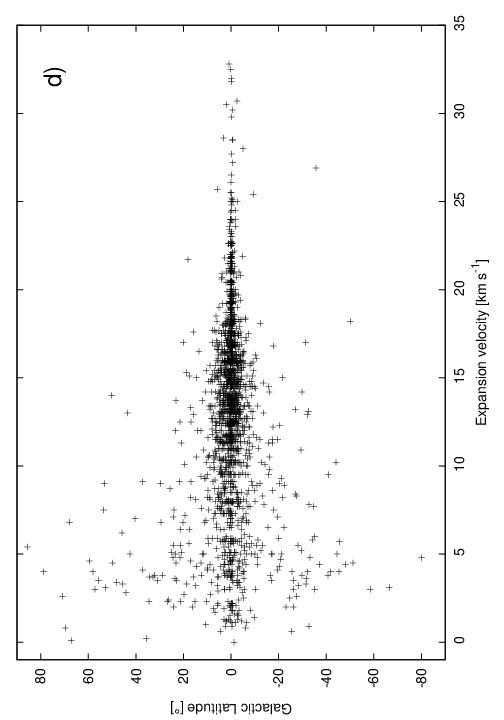}
  \end{flushright}
 \end{minipage}
\end{minipage}
   \caption{Statistics of 2341 detected OH masers in the database.
     The upper panels show in a) the distribution in Galactic
     longitudes, and in b) in Galactic latitudes. The lower panels
     show in c) the distribution on the sky in Galactic
     coordinates, and in d) the range of Galactic latitudes for given
     expansion velocities.  This last plot uses velocity information
     from 1967 stars and is truncated at 35 \kms\ for display
     purposes, excluding 15 sources with $v_{exp} = 35-80$ \kms.
              \label{statistics}%
           }
\end{figure*}

Surveys for OH maser emission were mostly made at 1612 MHz and only
occasionally in the weaker main lines. The major surveys, which
discovered the bulk of the 1612 MHz OH masers were the {\em ATCA/VLA
  OH 1612 MHz survey} by Sevenster et al. (\cite{Sevenster01}) with
766 detections, the {\em Dwingeloo/Effelsberg/Parkes 1612 MHz OH
  survey of IRAS point sources} by Te Lintel Hekkert et
al. (\cite{TeLintel91}) with N$\approx$740 detections, and the {\em
  Arecibo 1612 MHz survey of color-selected IRAS sources} by
B.M. Lewis and collaborators (Lewis \cite{Lewis94} and references
therein) with N$\approx$400 detections.

Most of the main-line observations are follow-up observations of stars
already detected at 1612 MHz. The detection rate for 1667 (1665) MHz
masers among stars with 1612 MHz detections is 62\% (39\%). The
detection rate for 1665 MHz masers raises to 58\%, if 1612 and 1667
masers are both present. Among stars observed but not detected at
1612 MHz the detection rates for 1667 and 1665 MHz masers decrease to
10\% and 9\%. Rarely a 1665 MHz maser is detected without accompanying
1667 MHz emission. Among the stars not detected at 1612 and 1667 MHz
the detection rate for 1665 MHz masers is only 1.5\%.

The numbers given here are affected by selection effects which are
difficult to quantify, because of the heterogeneous surveys and
follow-up observations contributing to the database. However, a
comparison with the major targeted surveys suggests that the numbers
are representative.  The {\em Dwingeloo/Effelsberg/Parkes} survey made
at 1612 MHz toward IRAS sources with flux densities $S_{12\mu m}>3$
Jy had a detection rate of $\approx 30$\% (Te Lintel Hekkert et
al. \cite{TeLintel91}). The {\em Arecibo 1612 MHz survey} has a
superior sensitivity and increased the detection rate among IRAS sources
with flux densities $S_{25\mu m}>2$ Jy to 39\% (Lewis
\cite{Lewis94}). Both surveys used IRAS color selected samples and
make up for 34\% of all observations contained in the database. Their
rates are close to 35\%, the fraction of detected 1612 MHz masers
among all sources listed in the database.  For the main lines, Lewis
(\cite{Lewis97}) obtained 1667 and 1665 MHz detection rates of 56\%
and 35\% among IRAS sources previously detected at 1612 MHz.  Lewis'
observations were made with a sensitivity of 20 mJy (3$\sigma$). Both
rates are quite similar to the detection rates given above for the
database. Lower rates were reported by David et al. (\cite{David93b}),
who obtained a detection rate at 1667 MHz of 24\% among IRAS color
selected sources.  For sources also detected at 1612 MHz this rate
increases to 31\% (Le Squeren et al. \cite{LeSqueren92}). Their lower
rates are likely caused by the poorer sensitivity limit of 120 mJy
(3$\sigma$). The role of the sensitivity of the observations is also
evident, if the main line detection rate is calculated in the database
as a function of the 1612 MHz flux density.  Dividing the sample of
sources with 1612 MHz detections according to peak flux density F(1612
MHz) into two parts equal by number, the 1667 MHz detection rate is
67\% for F(1612 MHz) $> 0.5$ Jy and 44\% below this level.

\subsection{Flux densities and shell expansion velocities \label{flux}}
The database provides radial velocities $v_{blue}$, $v_{red}$ and peak
flux densities $f_{blue}$, $f_{red}$ of the masers for detected
stars. For stars with OH masers exhibiting a double-peaked profile
the radial velocity $v_{rad}$ of the star and the expansion velocity
$v_{exp}$ of the shell can be derived from the peak velocities.
Figure \ref{velexp-flux} shows the distribution of the 1612 MHz peak
flux densities of the masers and of the shell expansion velocities up
to $v_{exp}=40$ \kms.  For the construction of the flux density
distribution the strongest of the maser emission features is used. The
distribution peaks at $\log f_{peak} \approx -0.3\pm0.2$ ($f_{peak}
\approx 300-800$ mJy), which is close to the sensitivity limit of the
major surveys. The decrease of the distribution below this peak is
therefore likely due to incompleteness of the database.

The distribution of shell expansion velocities is composed of three
parts. Most of the sources (68\%) have $10 \le v_{exp} \le 20$
\kms\ with a pronounced peak at 14 \kms. A quarter of the sources
(24\%) make up the low-expansion-velocity part, of which about 47\%
have Galactic latitudes $b>\; \mid$$5^\circ$$\mid$ (see below). A minority
(7\%) have unusually high expansion velocities with $v_{exp} > 20$
\kms. The extreme sources with $v_{exp} > 35$ are M-Supergiants,
post-AGB stars, unclassified objects, and two cases where the
maser peak identifications are doubtful.

\subsection{Distribution in Galactic coordinates}
The distribution of the stars exhibiting OH maser emission in Galactic
coordinates is shown in Fig. \ref{statistics}. The distribution in
longitude peaks in the Galactic Center region ($l \approx 0^{\circ}$).
Outside the longitude range $-90 < l < +90^{\circ}$ shown, 247 stars
(10.6\%) were detected. The lower detection rate at larger longitudes
was found already by the early surveys (Le Squeren et
al. \cite{LeSqueren92}; Lewis \cite{Lewis94}). Lewis attributes this
difference to variations in the ambient UV field, which is responsible
for the production of OH via the dissociation of \water\ molecules.
There is a pronounced asymmetry in the number of OH emitting stars
with respect to the Galactic Center (Fig. \ref{statistics}a and c). At
$+10 < l \le +90^{\circ}$ the number of detected stars is a factor of 1.8
higher than in the corresponding range at $-90 < l \le
-10^{\circ}$. This is most likely a selection effect due to the
contributions of the larger radio telescopes in the northern
hemisphere on Earth. However, Sevenster et al. (\cite{Sevenster01})
found in their homogeneous {\em ATCA/VLA OH 1612 MHz survey} a similar
but less pronounced asymmetry. They attributed the higher incidence of
detections by a factor of 1.25 to the presence of the Galactic bar
extending to higher longitudes on the northern side of the Galactic
Center.

The   distribution of the stars in Galactic latitude are concentrated
toward the Galactic plane (Fig. \ref{statistics}b). Outside the
latitude range $-15 < b < +15^{\circ}$ shown, only 198 stars (8.5\%)
were detected.  This concentration toward the Galactic plane is in part
due to the restricted latitude coverage of the major OH maser surveys.
The lower panels in Fig. \ref{statistics} show the Galactic
distribution in a different manner and show the dependence of the
expansion velocities on Galactic latitude. The pronounced peak at the
Galactic Center region in both longitude and latitude
(Fig. \ref{statistics}a-c) is biased owing to several dedicated surveys
in this region. In Fig. \ref{statistics}d, the larger spread in
Galactic latitude at small expansion velocities ($v_{exp} \le 10$
\kms) is  real, however,  and is explained by Lewis (\cite{Lewis91}) as a
superposition of low-latitude high-mass AGB stars with $v_{exp}
\approx 15$ \kms\ and low-mass AGB stars with $v_{exp} \approx 5$
\kms\ distributed over a much larger latitude range.  With few
exceptions, high expansion velocities (v$_{exp}>$20 \kms) are found
only among stars in the Galactic plane ($\mid$b$\mid <2^\circ$),
implying that v$_{exp}$ is statistically an indicator for
main-sequence mass. Higher expansion velocities are reached only by
the higher mass AGB stars.

\subsection{OH maser luminosities}
To characterize the observed strength of a maser,  peak flux
densities are usually given in the literature. Integrated fluxes are only
rarely available, and therefore we calculated OH maser
luminosities as specific luminosities $L_\nu = f_{peak} * 4 \pi D^2$
using peak flux densities $f_{peak}$, as derived in Sect. \ref{flux},
and adopting kinematic distances $D$.  To calculate kinematic
distances we used the prescription from Reid et al. (\cite{Reid09})
with updated Galactic and solar motion parameters. The distance to the
Galactic Center was set to $R_0 = 8.4$ kpc and the circular rotation
speed was set to $\Theta_0 = 247$ \kms\ (Brunthaler et
al. \cite{Brunthaler11}). At 1612 MHz only sources with a double-peaked maser
spectrum were used because otherwise the radial velocity is rather
uncertain. In those cases, where the kinematic distance is ambiguous,
we used the near kinematic distance. For the main lines we used only
sources that have a near kinematic distance determined with 1612 MHz
velocity information.

\begin{figure}
   \includegraphics[width=6.0cm,angle=-90]{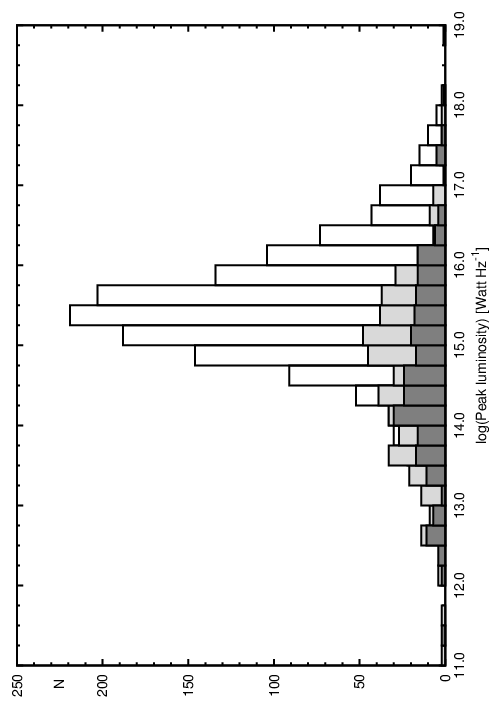}
 \caption{OH Maser specific luminosity distributions $L_\nu$ in
     Watt Hz$^{-1}$ assuming near kinematic distances.  Included are
     1456 sources with reliable radial velocities, e.g. having
     double-peaked spectra at 1612 MHz (unfilled bars). Superposed are
     the distributions of 461 sources with 1667 MHz emission (bars
     with light gray shading) and  275 sources with 1665 MHz
     emission (bars with heavy gray shading).}
 \label{maser_lumino}%
\end{figure}

The distribution of the maser specific luminosities $L_\nu$ is shown
in Fig. \ref{maser_lumino}. The core of the 1612 MHz OH maser
luminosity distribution can be modeled by a Gaussian curve centered on
$\log (L_\nu) = 15.30$ and with a width FWHM = 1.25, corresponding to
a typical luminosity range $0.5 - 8.4 \times 10^{15}$ Watt
Hz$^{-1}$. In principle, the distribution displayed will underestimate
the true luminosity distribution because part of the sources will be
located at larger distances corresponding to the far kinematic
distances. Their luminosities are underestimated, but there is no
distinct excess feature in the low-luminosity tail of the
distribution, which could be assigned to such sources. A recalculation
of the specific luminosity distribution using far kinematic distances
leaves the width unchanged and shifts its center to $\log (L_\nu) =
15.50$. The mean specific luminosity $<$$L_\nu$$>$ at 1612 MHz is
therefore between 2 and $3 \times 10^{15}$ Watt Hz$^{-1}$.

The distributions in specific luminosities $L_\nu$ of the main lines
are rather broad and similar to each other. They cannot be represented
by a Gaussian curve and therefore we determined the mean specific
luminosities using the median.  This gives $<$$\log (L_\nu)$$>_{1667
  MHz} = 14.75$ and $<$$\log (L_\nu)$$>_{1665 MHz} = 14.59$ compared
to $<$$\log (L_\nu)$$>_{1612 MHz} = 15.42$. The main lines are
therefore a factor $\approx$6 less luminous than the satellite 1612
MHz line.  However, the ratio of specific luminosities between the
1612 MHz line and the main lines is dependent on the mass-loss
rates. At lower rates typical for Mira variables this ratio is small,
so that often main lines are detected while the 1612 MHz line is not.
At higher rates typical for OH/IR stars the 1612 MHz line clearly
prevails (Lewis \cite{Lewis89}). Therefore the mean specific
luminosities give the gross average for all sources in the database
only.

\section{Discussion}
The generally broad distributions in the main parameters
characterizing the OH maser emission (velocities, specific
luminosities) and their host stars (Fig. \ref{statistics}) indicate
that these masers arise from  diverse  environments. The
characterization of these different environments requires the analysis
of the optical and infrared emission of the stars, which will not be
followed here. Such comprehensive and still up-to-date analyses are
provided by Chen et al. (\cite{Chen01}) and Kwon \& Suh
(\cite{Kwon12}).  The major contributors to the database are AGB stars
with oxygen-rich chemistry. This includes Mira variables, OH/IR stars,
and probably a sizeable fraction of post-AGB stars. The masers are
located in spherical circumstellar dust shells with different dust
densities and expansion velocities. Among the post-AGB stars the
masers may arise also in dusty disks or tori.  Minor contributors are
M-supergiants, planetary nebulae, and a few silicate-carbon stars.
While the low number of M-supergiants is likely due to the low number
of M-supergiants in general, the OH emitting planetary nebulae and
silicate-carbon stars are peculiar within their parent population by
having non-standard environments. In planetary nebulae amplification
of the radio continuum emission might be held reponsible (Uscanga et
al. \cite{Uscanga12}), while the OH masers in silicate-carbon stars
may arise like the \water\ masers in oxygen-rich disks associated with
binary systems containing a carbon star (Szczerba et
al. \cite{Szczerba06}).

The database provides easy access to the information scattered in the
literature about Galactic OH masers in the CSEs of late-type stars.
This will help to identify efficiently known masers in future surveys,
which are planned within the frame of the SKA and its precursor
instruments. The Galactic ASKAP survey alone expects to discover
several thousand new masers in the Galactic plane and the Magellanic
Clouds (Dickey et al. \cite{Dickey13}). With SKA, the search for
stellar masers will be extended to local group galaxies, and
comparisons to the maser population in the Milky Way will then take
advantage of the database (Etoka et al. \cite{Etoka14}).  The database
will contribute to satisfying the need for an efficient access to the
accumulated knowledge about known masers in the cm wavelength range,
which is likely to increase with the rising activities in millimeter
and submillimeter maser research triggered by ALMA.

\begin{acknowledgements}
We thank B. Heidmann and C. Matthies for their support with the
maintenance of the database, M. Demleitner for the setup of the VO
service, and S. Etoka and the referee for comments on the
manuscript. The database and the associated software was constructed
as part of the activities of the German Astrophysical Virtual
Observatory (GAVO). This work is supported by the Deutsche
Forschungsgemeinschaft under grant number \mbox{En~176/33}.
\end{acknowledgements}

\end{document}